\documentclass[pre,aps,
twocolumn,
superscriptaddress,floats,showpacs,10pt]{revtex4-1}
\usepackage{graphicx}
\usepackage{dcolumn}
\usepackage{bm}
\usepackage{amssymb}
\usepackage{amsmath}
\usepackage{array}
\usepackage{float}
\usepackage{geometry}
\usepackage{color}
\usepackage{amsthm}

\begin{document}

\title{Controlling the generation of high frequency electromagnetic pulses with relativistic flying mirrors using an inhomogeneous plasma}
\author{Mathieu Lobet}
\affiliation{ENSEIRB-MATMECA, University of Bordeaux, 1 Av. Du Dr. Schweitzer, Talence, 33400 France}
\affiliation{Kansai Photon Science Institute, JAEA, Kizugawa, Kyoto 619-0215, Japan}
\author{Masaki Kando}
\affiliation{Kansai Photon Science Institute, JAEA, Kizugawa, Kyoto 619-0215, Japan}
\author{James K. Koga}
\affiliation{Kansai Photon Science Institute, JAEA, Kizugawa, Kyoto 619-0215, Japan}
\author{Timur Zh. Esirkepov}
\affiliation{Kansai Photon Science Institute, JAEA, Kizugawa, Kyoto 619-0215, Japan}
\author{Tatsufumi Nakamura}
\affiliation{Kansai Photon Science Institute, JAEA, Kizugawa, Kyoto 619-0215, Japan}
\author{Alexander S. Pirozhkov}
\affiliation{Kansai Photon Science Institute, JAEA, Kizugawa, Kyoto 619-0215, Japan}
\author{Sergei V. Bulanov}
\altaffiliation[Also at ]{Prokhorov Institute of General Physics, Russian Academy of Sciences, Moscow
119991, Russia}
\affiliation{Kansai Photon Science Institute, JAEA, Kizugawa, Kyoto 619-0215, Japan}

\date{18.09.2012}

\begin{abstract}

A method for the controlled generation of intense high frequency electromagnetic fields by a breaking Langmuir wave 
(relativistic \textit{flying mirrors}) in a gradually inhomogeneous plasma is proposed. 
The wave breaking threshold depends on the local plasma density gradient. Compression, chirping and 
frequency multiplication of an electromagnetic wave reflected from relativistic mirrors is demonstrated using 
Particle-In-Cell simulations. Adjusting the shape of the density profile enables control of the reflected light properties.

\end{abstract}

\maketitle


%

High frequency coherent radiation sources are of great interest for various applications and future understanding in fundamental 
science \cite{Mourou2006, Krausz2009}. In material science, the study of extremely short time reactions requires high temporal and 
spatial resolution imaging which could be achieved using extremely short duration high frequency electromagnetic pulses. 
High-power X-ray radiation can be used to analyse nano-scale objects such as complex molecules offering low noise and sufficient 
energy before their destruction due to Coulomb repulsion \cite{Neutze2000}. In laser-matter interaction, the possibility to develop 
high frequency extremely-intense laser systems would lead to novel quantum electrodynamics effects \cite{QED}.

An approach to generate high frequency radiation based on the concept of the \textit{flying mirror} 
considers a plasma shell travelling close to the speed of light as a relativistic mirror. Reflected light undergoes double 
Doppler frequency up-shift, compression, intensification and focusing due to relativistic effects. 
Various schemes were described \cite{Bulanov2003, Bulanov2006, Kulagin2007} and experimentally demonstrated 
\cite{Kando2007-2009} as a proof of the feasibility of this concept. 

Controlling the generation time and position of the breaking wave is an essential aspect for experiments 
to synchronize the mirrors with an incident laser pulse in an accurate, flexible and facilitated way. 
It also allows for tuning the reflected pulse properties.

In the laser wakefield acceleration regime \cite{Tajima1979}, wave breaking offers electrons the initial injection 
required to enter the acceleration phase of the wakefield \cite{Bulanov1992}. For this purpose, tailored inhomogeneous 
plasmas have been considered as a way to control the wave breaking. In a downward density profile, the plasma frequency 
depends on the coordinates. In the wave, electrons progressively oscillate out of phase resulting in the breaking of the wave 
\cite{Dawson1959}. This mechanism was examined in Refs. \cite{Bulanov1998,Brantov2008,Geddes2008} in the case of an inhomogeneous 
plasma with scale length larger than the plasma wavelength. Step-like density transitions are also useful for controlled electron 
injection \cite{Suk2001,Brantov2008}. They can be produced experimentally using a transverse laser pulse \cite{Chien1995} or using 
an obstacle (razor blade) inserted into the ultrasonic gas flow \cite{Schmid2010}. Recent experiments in  Refs. \cite{Faure2010} 
use a density depletion channel generated in a gas-jet by a transverse laser pulse.  

Under the realistic conditions of experiments, the plasma is always inhomogeneous. In the present paper, 
we examine plasma inhomogeneity effects on the \textit{flying mirrors} and propose using plasma inhomogeneity 
as a new way to control the generation of localized \textit{flying mirrors}. We use 1D Particle-In-Cell (PIC) 
simulation for detailed analysis of the formation of the mirror and the reflection of the light. The possibility 
to tune the properties of the reflected light by adjusting the plasma parameters is demonstrated.



An electromagnetic pulse travelling in an underdense plasma (i.e. the local density $n$ is below the critical density $n_c = m_e \omega^{2} /4 \pi e^2$ where $\omega$ 
is the laser frequency, $m_e$ is the mass of an electron and $e$ its charge) generates a driven electron plasma wave. 
In the linear regime, the wave has a sinusoidal shape and electrons oscillate at 
the non-relativistic plasma frequency $\omega_{p} = \sqrt{4\pi n_e e^2/m_e}$ where  $n_e$ is the local electron density. 
The plasma wave propagates with phase velocity close to the group velocity of the laser driver pulse $ v_{ph} \simeq v_g$ \cite{Tajima1979}.

For a sufficiently intense driver pulse, one enters the nonlinear regime. The plasma wave breaks when 
$v_{ph} \leq v_{e}$ with $v_e$ being the electron velocity. Nonlinear effects cause the steepening of the wave. 
The electron density appears strongly modulated with high density shells formed at the positions of maximum velocity \cite{Akhiezer1956}.

In an underdense plasma, the Lorentz factor calculated for $v_e$, $\gamma _e =1/\sqrt{1-(v_e/c)^2}$, 
can be approximated by $\gamma _e = 1 + a^2/2$ assuming a stationary wave solution, where $a = eE/mc \omega$ 
is the normalized laser amplitude and $E$ is the laser electric field. The Lorentz factor for the phase velocity, $\gamma_{ph} =1/\sqrt{1 - (v_{ph}/c)^2}$, 
is related to the plasma frequency as $\gamma_{ph} = (\omega/\omega_p)\sqrt[4]{1+a^2}$, 
where the dependence of the laser group velocity $v_g$ on the laser amplitude has been taken into account. 
Writing the breaking condition $\gamma _{e} = \gamma _{ph}$ \cite{Akhiezer1956} in the form $a^2/2 + 1= \omega / \omega _p \sqrt[4]{1+a^2}$, 
we obtain \cite{Zhidkov2004} the breaking threshold
\begin{eqnarray}
a_{br} &= \left \{
	\begin{array}{rcr}
		 \sqrt{ 2 \left( {\omega }/{\omega _{p}} - 1 \right)} &\quad {\rm for} \quad a \leq 1 \\
		 (2\omega/\omega_p)^{2/3} &\quad {\rm for} \quad a \gg 1 \\
	\end{array}
	\right.
.
\label{eq:breaking_condition}
\end{eqnarray}

Strong density modulations formed in the wake of an intense driver pulse constitute relativistic 
\textit{flying mirrors}. A counter-propagating electromagnetic pulse, referred to as the source pulse, 
is partially reflected back with frequency up-shifting and compression. For normal incidence reflection, 
the reflected pulse frequency $\omega _{r}$ is given by the relationship
\begin{equation}
\omega _{r} = \omega _{s} \frac{1+\beta _{ph}(t)}{1-\beta _{ph}(t)}, \label{eq:doppler_effect}
\end{equation}
for $\beta_{ph}(t) = v_{ph}(t)/c$ where changes in $\beta_{ph}(t)$ are much slower than $\omega_s^{-1}$. 
In the case of a relativistic mirror $\beta_{ph} \rightarrow 1$, Eq. (\ref{eq:doppler_effect}) 
can be approximated by $\omega _r \simeq 4 \gamma _{ph}^2 \omega _s$. 
The reflected electromagnetic field $E_r$ is amplified by the same factor so that $E_r = E_s (1+\beta _{ph}(t))/(1-\beta _{ph}(t))$. 
In the general case, the dependence on time of $\beta _{ph}$ results in the chirp of the reflected light \cite{Esirkepov2009}.

The reflected pulse energy is determined by the mirror reflectivity. 
For a small amplitude plasma wave, the reflectivity is exponentially small. Close to the wave breaking, 
the electron density becomes singular providing high reflectivity with the number of reflected photons being proportional to $1/\gamma _{ph}^4$ \cite{Panchenko2008}. 


\begin{figure}[t]
\includegraphics[width=7.5cm]{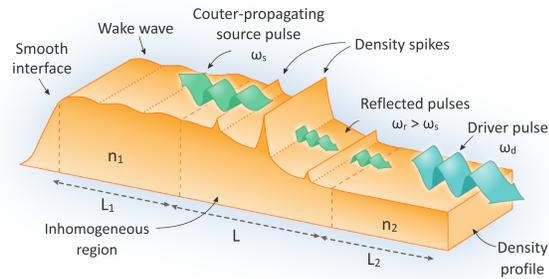}
\caption{
Tailored inhomogeneous density profile in the framework of the \textit{flying mirror}.
}
\label{fig:inh_density_profile}
\end{figure}

In an inhomogeneous plasma, the plasma wave becomes a wave with a continuous spectrum \cite{Dawson1959, Bulanov1998}. 
Its wave number, $k_p$, depends on time $t$ and position $x$ according to the equation
\begin{equation}
\partial _t k_p = -\partial _x \omega _{p}.
\end{equation}
This equation is a consequence of the relationships $\omega = -\partial _t \theta$, $k = \partial _x \theta$ 
and the cross differentiation property $\partial _{tx} \theta = \partial _{xt} \theta$ where $\theta(x,t)$ is the eikonal of the plasma wave.
It yields $k_p = k_{p,0} - \partial _x \omega _p t$. The phase velocity, equal to $v_{ph} = \omega _p / k_p$, therefore is given by
\begin{equation}
v_{ph} = \frac{\omega _{p}}{ k_{p,0} - \partial_x \omega _{p} t} \label{eq:inh_vph},
\end{equation}
where $k_{p,0}$ is the initial wave number. The plasma frequency $\omega _p$ 
depends on the local value of the electron density $n_e(x)$ which yields $\partial _x \omega _{p} = \sqrt{\pi e^2 /m_e}\partial _x n_e/ \sqrt{n_e}$. 
In a gradual inhomogeneous density profile $\vert \partial _x \omega _{p} \vert \ll 1$ we have
\begin{equation}
v_{ph} \simeq  v_{ph,0}\left(1 + \sqrt{\frac{\pi e^2}{m_e}} \frac{\partial _x n_e}{n_e} t \right)
\label{eq:approx_vph}
\end{equation}
where $v_{ph,0}$ is the phase velocity in a homogeneous plasma.
In a downward density profile $\partial _x n_e < 0$, the phase velocity progressively decreases with time. 
The corresponding deceleration depends on the magnitude of the density gradient $\vert \partial_x n_e \vert$. 
The breaking condition $\gamma_{ph} = \gamma _{e}$ is inevitably reached after some wave periods. 
The breaking time $t_{br}$ \cite{Brantov2008} is equal to
\begin{equation}
t_{br} = \frac{ 2 n_e }{c\vert \partial _{x} n _{e} \vert}\left( \beta_{e}^{-1} - \beta_{ph}^{-1} \right).
\label{eq:t_br_inh}
\end{equation}

In order to study the flying mirror properties in inhomogeneous plasmas we perform numerical simulations using a $\textrm{1D 2/2}$ parallel PIC code. 

Fig. \ref{fig:inh_density_profile} illustrates the assumed tailored density profile composed 
of two homogeneous plasma slabs of density $n_1 = 0.025 n_c$ ($\simeq 0.275\times 10^{20}\ \mathrm{cm}^{-3}\times (1 \mu \mathrm{m}/\lambda)^2$) 
and $n_2 = 0.0175n_c$ corresponding to plasma wavelengths of $\lambda_{pe,1} \simeq 40 \lambdabar$ and $\lambda_{pe,2} \simeq 47.5 \lambdabar$ 
and respective lengths $L_1 = 200 \lambdabar$ and $L_2 = 350 \lambdabar$ where $\lambdabar = c\omega^{-1}$. 
They are separated by a decreasing inhomogeneous region having a cosine square density profile, $n_e(x) = (n_1 - n_2)\cos^2 (\pi x / L) + n_2$, 
which provides continuous density variation from $n_1$ to $n_2$ with the length $L = 150 \lambdabar$ ($\simeq 24\ \mu \mathrm{m}$ for $\lambda = 1\ \mu \mathrm{m}$) 
which approximately corresponds to three plasma wavelengths. To limit the effect of the \textit{vacuum heating} \cite{DI1979}, 
a pre-plasma of length $L_0 = 200c\omega^{-1}$ gradually rises from zero to $n_1$ at the left-side of vacuum-plasma interface.

The driver pulse is Gaussian linearly polarized along the $\mathbf{y}$ direction with wavelength $\lambda _d = \lambda$, 
longitudinal full width at half maximum (FWHM) $\textrm{l} _d =10 \lambda $ and its amplitude is equal to
$a_d = 1.5$ ($I \simeq 3.1\times 10^{18}\ \mathrm{Wcm}^{-2} \times (1\ \mu \mathrm{m} / \lambda)^2$). 
Densities in the homogeneous regions give for the breaking condition $a_{br} = 3.2$, from Eq. \ref{eq:breaking_condition}, 
so that the generated wake wave is initially below the breaking level $a_d < a_{br}$. 
The source pulse is rectangular and is polarized in the direction perpendicular to the driver 
polarization. Its length is equal to $l_s = 80 \lambda$, $\lambda _s = 2\lambda $ and the amplitude $a_s = 2 \times 10^{-4}$ 
is  relatively low to not disturb the driven plasma wave.
\begin{figure}[t]
	\includegraphics[width=7.5cm,height=13.75cm]{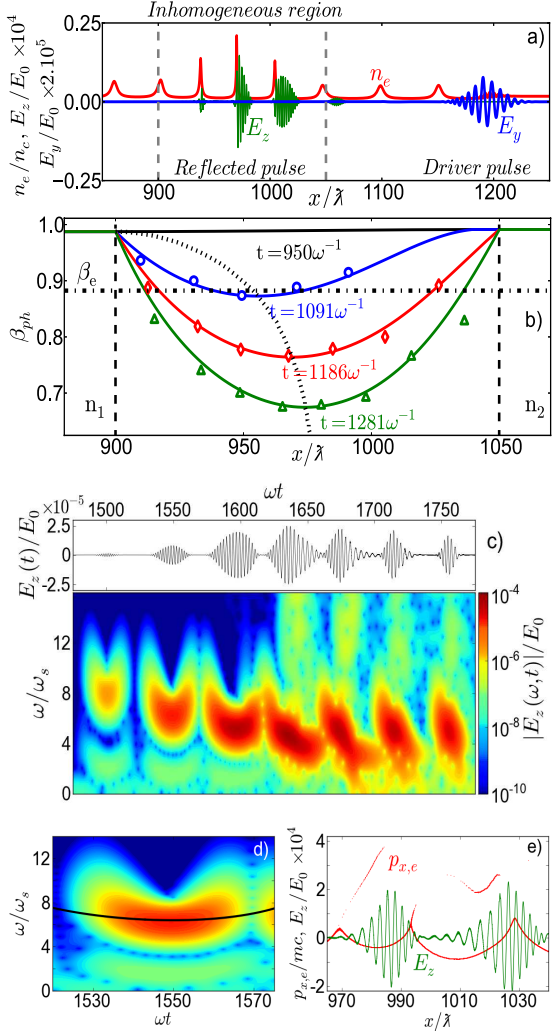}
\caption{
Results of the PIC simulation: a) Generation of the density spikes in the plasma wave ($n_e$) 
and reflection of the source pulse ($E_z$). b) Evolution of the plasma wave phase velocity $\beta _{ph}$ 
in the inhomogeneous plasma region. 
Solid curves correspond to the theoretically calculated $\beta _{ph}$, markers give $\beta _{ph}$ found 
in the simulation and the dotted line indicates the position of the minimal $\beta _{ph}$. c) 
Time-frequency spectrum $\vert E_z(\omega,t) \vert$ of the electric field $E_z(\omega)$ 
from the light reflected by the mirrors. d) Analysis of a parabolic chirp. 
The line corresponds to the analytical evolution of the frequency. e) Acceleration of electrons by the breaking plasma wave. 
A low pass filter was used to extract the electric field of the reflected light from the main source pulse. 
}
\label{fig:simulation_1}
\end{figure}

In Fig. \ref{fig:simulation_1}a, the simulation results are presented. The driver pulse propagating in the positive $\mathbf{x}$ 
direction initially generates a weakly nonlinear plasma wave. The wave only breaks in the inhomogeneous region. 
Density spikes develop in a finite time and constitute temporary existing mirrors.  Since we know the breaking time (\ref{eq:t_br_inh}), 
this process ensures spatial and temporal control of the breaking wave. 

The source pulse is longer than the plasma wavelength, the reflection on the periodic train of density spikes generates a complex signal composed of several wave packets.

The time dependence of the wake wave phase velocity versus position in the inhomogeneity is shown in Fig. \ref{fig:simulation_1}b for different times. 
The numerical results (shown by markers) appear in good agreement with the theoretically decreasing evolution of the phase velocity (solid lines) 
given by Eq (\ref{eq:inh_vph}). The breaking condition is at first satisfied at $t = 130 \omega^{-1}$ after the driver pulse has entered the target. 
The position of the minimal phase velocity progressively tends to the position of the maximal density gradient.

The time-frequency spectrum of the reflected radiation is presented in Fig. \ref{fig:simulation_1}c where the electric field $E_z$ 
is plotted in the plane $(\omega,t)$. The time-frequency analysis of $E_z(t)$ is performed using the Gabor transformation defined as \cite{FS2002}
\begin{equation}
E_z(\omega,t) = \int_{-\infty}^{+\infty}{E_z(\xi)e^{-i\omega \xi-\alpha(\xi - t)^2} d\xi}
\end{equation}
where $\alpha = 4 \log{(2)}c^2/l_w^2$ so that $l_w$ is the full length at half maximum of the Gaussian window.

A remarkable property of the high frequency radiation is the generation of a train of short electromagnetic wave packets. 
Space and time evolution of the mirror velocity, $v_{ph}$, leads to the generation of parabolic chirps as shown in Fig. \ref{fig:simulation_1}d. 

In order to calculate the frequency chirp we consider a plasma frequency, in the inhomogeneous region, of the form
\begin{equation}
\omega_p(x) = \frac{\Delta \omega}{2} \tanh\left(\frac{x}{L}\right) + \frac{\omega _1 + \omega _2}{2} 
\end{equation} 
with $\Delta \omega = \omega _1 -\omega _2  $. From Eq. (\ref{eq:inh_vph}), the phase velocity can be locally approximated by
\begin{equation}
v_{ph} = v_{ph,min} + \delta x^2,
\end{equation}
around the position of minimal phase velocity $v_{ph,min}$ with $\delta = \Delta \omega t / 2 L^3 k_p$. Using Eq. (\ref{eq:doppler_effect}), 
we obtain for the frequency multiplication the following quadratic equation
\begin{equation}
\omega _r = \omega _{r,min} + \frac{\delta}{1 - \beta _{ph,min}}x^2
\label{eq:frequency_multiplication}
\end{equation}
where $\omega _{r,min}$ is the frequency multiplication after reflection on a mirror travelling at $v_{ph,min}$. Eq. (\ref{eq:frequency_multiplication}) 
characterises the parabolic frequency modulation of the reflected pulses.

When electrons are injected into the wakefield, the mirrors decelerate due to momentum conservation which causes the wave packets to be down chirped. 
This phenomenon happens at the reflection from the fourth mirror formed behind the driver pulse as shown in Fig. \ref{fig:simulation_1}e.
Due to the decreasing mirror velocity $v_{ph}$, density maxima gradually increase offering higher reflectivity. The life spans of the mirrors are 
gradually prolonged. The average reflected frequency diminishes and the spectrum broadens according to Eq. (\ref{eq:doppler_effect}).

The properties of the reflected pulse can be modified by changing the parameters of the inhomogeneous region: density profile, length $L$ 
and density variation $\Delta n_e = n_1 - n_2$.


\begin{figure}[h!]
\includegraphics[width=7.5cm]{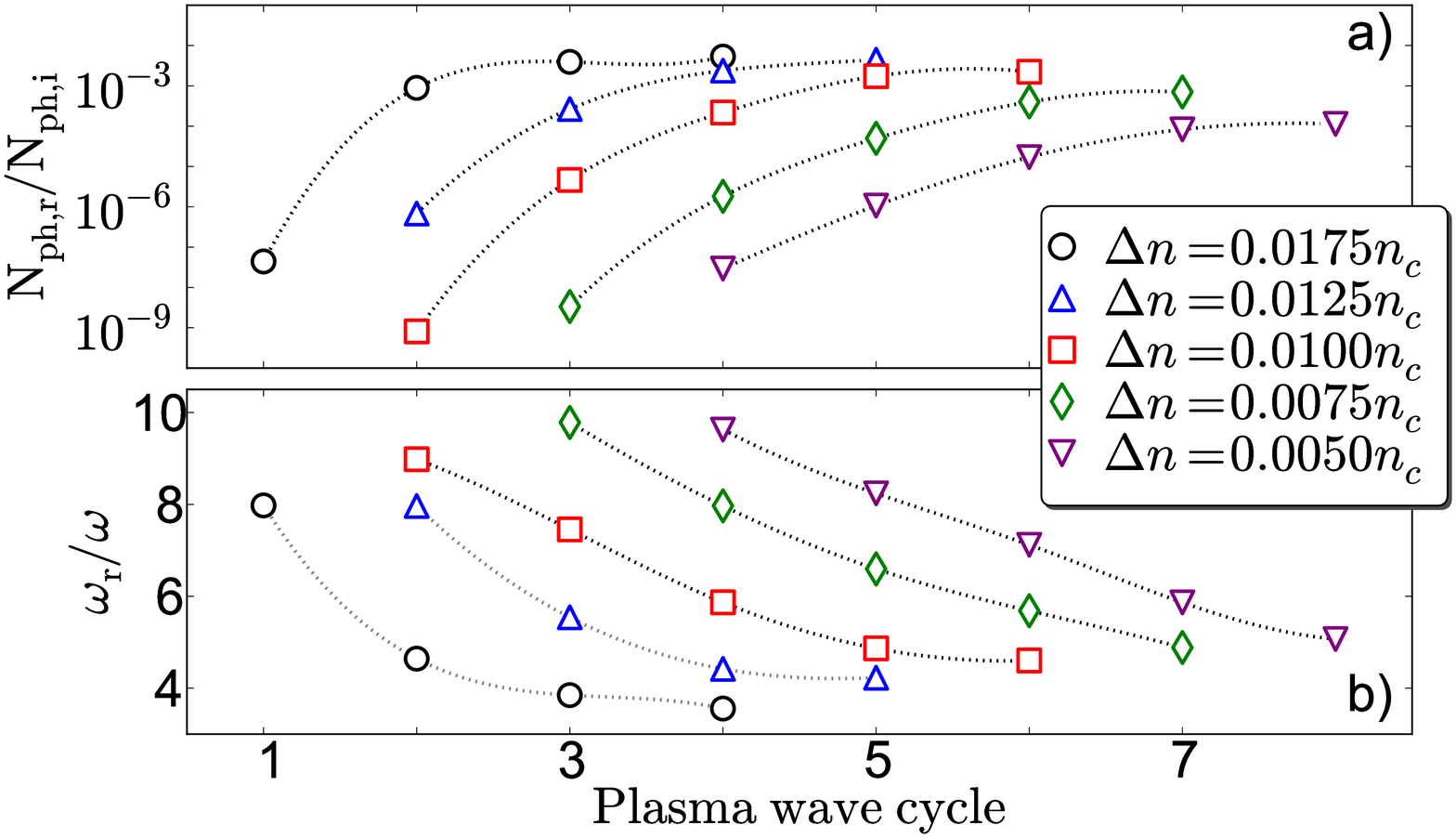}
\caption{
a) Number of reflected photons normalized by the number of incident ones $N_{ph,r}/N_{ph,i}$. b) Average reflected frequency for density 
variation $\Delta n_e$ changing from $0.005n_c$ to $0.0175n_c$ as a function of the plasma wave cycles at which the reflections have occurred.
}
\label{fig:dependence_parameters}
\end{figure}

\begin{figure}[h!]
\includegraphics[width=7.5cm]{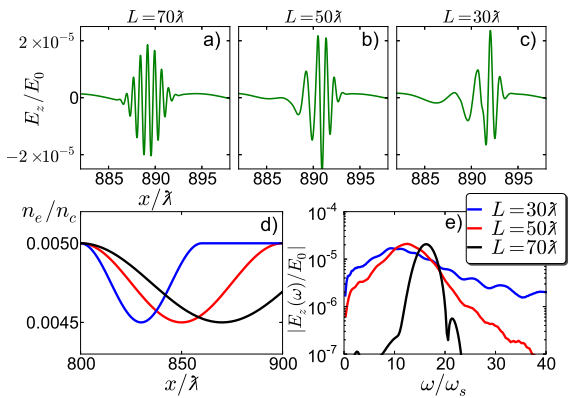}
\caption{
a), b), c) Normalized density $n_e$ and electric field $E_z$ at time $t = 1100 \omega^{-1}$ for $L = 70$, $50$ and $30\lambdabar$. 
d) Initial normalized density pit $n_e$. e) Spectra $\vert E_z(\omega) \vert$ of the reflected wave packets for the different pit lengths.
}
\label{fig:phase_sharp}
\end{figure}

In the spectral domain, average reflected frequency can be defined as
\begin{equation}
\bar{\omega} = \frac{\displaystyle{\int_{0}^{+\infty}{\vert E(\omega) \vert} \omega d\omega}}{\displaystyle{\int_{0}^{+\infty}{\vert E(\omega) \vert } d\omega}}
\label{eq:average_frequency}
\end{equation}
and the number of photons as
\begin{eqnarray}
N_{ph} = \int_{0}^{+\infty}{\frac{E(\omega)^2 + B(\omega)^2}{8\pi^2 \hbar \omega} d\omega}
\label{eq:number_photon}
\end{eqnarray}
For density variation $\Delta n_e$ changing from $0.005n_c$ to $0.0175n_c$, the evolution of the reflected number of photons divided by the numbers 
of incident ones is shown in Fig. \ref{fig:dependence_parameters}a as a function of the plasma wave cycle when reflection occurred. 
The average frequency, $\bar{\omega}_{r}$, is plotted in Fig. \ref{fig:dependence_parameters}b. 
The length of the inhomogeneous region is equal to $L = 200\lambdabar$. We interpret the increase of the reflected photon number as 
the onset of the wave breaking. As predicted by Eq. (\ref{eq:t_br_inh}), the larger is the density variation $\Delta n_e$, 
the shorter the breaking time. For $\Delta n_e = 0.005 n_c$, the wave breaks at $138 \omega^{-1}$ after the driver pulse has entered 
the inhomogeneity region, three wave periods behind the driver pulse. For $\Delta n_e = 0.020 n_c$, wave breaks during the first plasma wave cycle. 
For a given density, the frequency decreases while the number of reflected photons grows. Progressive decreasing of the phase velocity gives 
higher reflectivity but lower frequency up-shift (Eq. \ref{eq:doppler_effect}). 

Now, we consider a localized density depression, as realized experimentally in \cite{Faure2010}, 
with a scale length shorter than the plasma wavelength $L < \lambda_{p,1}$. The density depression profile 
is composed of a first decreasing plasma region of length $L$ followed by a symmetric increasing one.  
As illustrated in Fig. \ref{fig:phase_sharp}d, the density varies from $n_{e} = 0.005n_c$ to a minimal value $n_e = 0.0045 n_c$ ($\lambda_p \simeq 90 \lambdabar$) 
for lengths $L = 30$, $50$ and $70 \lambdabar$. The normalized amplitude of the driver pulse is equal to $a_d = 3$ whereas other parameters remain 
the same as previously described.

In the short-scale density perturbation, wave breaking leads to the generation of very short-lived mirrors. 
Reflected pulses have a very short duration ($\simeq 3\ \mathrm{fs}$ for $\lambda = 1\ \mu\mathrm{m}$), composed of a few cycles as 
displayed on Fig. \ref{fig:phase_sharp}a, b and c.
Decreasing the depression length reduces the reflection time and generates shorter wave packets. For the density depression lengths 
$L=30 \lambdabar$ (Fig. \ref{fig:phase_sharp}c) and $50 \lambdabar$ (Fig. \ref{fig:phase_sharp}b), the density gradient is sufficiently abrupt 
to result in the acceleration of an electron bunch seen in the density profile as a narrow density spike. This is of interest 
for the electron injection scheme \cite{Bulanov1998,Brantov2008,Geddes2008,Chien1995,Suk2001,Faure2010}. 
In this case, a characteristic down chirping of the reflected light can be observed. 
The spectra of the reflected signals in Fig. \ref{fig:phase_sharp}e show that the frequency was multiplied by a factor 
of 16 for length $L = 70 \lambdabar$. The shortening of the density depression length results in the generation of a stronger negative chirp.

In the downward density region, the plasma wave progressively travels with lower velocity than the reflected light. 
In the rising density part of the density depression, the plasma wave accelerates so that the distance 
to the reflected pulse can be reduced. In the simulations, the reflected pulse 
is not affected by the density modulation in the rising inhomogeneous density region.

In conclusion, a tailored downward plasma density offers time and space control for the generation 
of relativistic mirrors in a breaking laser-driven plasma wave. The reflected light properties 
including frequency multiplication, compression, chirping and duration can be modified by adjusting 
the density inhomogeneity parameters. This discovery is useful for experimental studies 
on the \textit{flying mirror} concept, including compression of controllably 
chirped attosecond pulses with aperiodic soft and hard X-ray mirrors \cite{Beigman2002}. 
It makes a new step in the realization of a stable, coherent, compact and tunable high frequency 
light source required for various applications. The analysis of the reflection of the laser source 
on a density profile can also be contemplated as a plasma diagnostic tool.

One of the authors, M. L., acknowledges the Region Aquitaine for grant Aquitaine Cap Mobilit\'e and expresses gratitude 
to P. Bolton and the Kansai Photon Science Institute for their generosity, their hospitality and their involvement in this project. 
The authors thank A. Ya. Faenov, T. A. Pikuz, Y. Hayashi, H. Kotaki and I. Daito for discussions.


\end{document}